# Boreal Afforestation's Underestimated Cloud Influence on Earth's Energy Imbalance

*Enoch Ofosu [a*], Kevin Bradley Dsouza [a], Daniel Chukwuemeka Amaogu [d], Jérôme Pigeon [d], Richard Boudreault [a,b,c,f], Juan Moreno-Cruz [d], Pooneh Maghoul [e,d], Yuri Leonenko [a]*

** Corresponding authors*

**Affiliations**

a - Department of Earth and Environmental Sciences, University of Waterloo
b - School of Environment, Enterprise and Development, University of Waterloo
c - AWN Nanotech
d - Department of Civil, Geological and Mining Engineering, Polytechnique Montréal
e - The United Nations University Institute for Water, Environment and Health (UNU-INWEH)
f - Université de Sherbrooke

## Abstract

Earth's Energy Imbalance (EEI) is accelerating, partly due to declining planetary albedo from reduced cloud cover. Boreal afforestation can either mitigate or exacerbate this trend through competing biophysical feedbacks. While snow masking reduces surface albedo (+0.5 to +2.5 W/m² warming), forests can enhance low-level cloud cover (0.1–0.5%) and increase cloud reflectivity via biogenic volatile organic compounds (BVOCs), producing potential cooling (–1.8 to –6.7 W/m²). This BVOC–aerosol–cloud pathway remains poorly constrained but may dominate under warmer conditions, challenging carbon-centric mitigation paradigms. Large-scale initiatives (e.g., Canada's *2 Billion Tree Commitment*) risk unintended warming if not climate-smartly sited. We present a tiered decision-support framework that integrates biogeochemical and biophysical processes, explicitly incorporating cloud–aerosol feedbacks. Urgent inclusion of these feedbacks in policy is essential to ensure boreal afforestation contributes to EEI stabilization.

## INTRODUCTION

The planet's energy ledger is tilting further out of balance. Earth's Energy Imbalance (EEI), the net difference between solar energy absorbed and thermal energy emitted has nearly doubled since the early 2000s, propelling the world toward record-breaking temperatures and crossing the +1.5 °C threshold in 2023 [1–3]. This surge is not only a story of greenhouse gas loading: satellite and surface records reveal a concurrent collapse in planetary albedo, driven in part by reductions in low-level cloud cover over northern mid-latitudes and the oceans [4–6]. Clouds, often treated as a modelling afterthought in land-based mitigation planning, are in fact pivotal regulators of Earth's radiative budget. Their sensitivity to land-cover change, particularly afforestation, represents one of the largest and least constrained feedbacks in the climate system.

Afforestation is embedded in the climate commitments of over 140 countries, including large-scale pledges such as Canada's 2 Billion Trees Program [7]. In the tropics, where high evapotranspiration, vigorous convection, and deep cloud systems dominate, the biogeochemical and biophysical effects of afforestation generally align to produce robust net cooling [8–10]. The boreal biome tells a more ambiguous story. Spanning ~30 % of global forest cover and holding vast carbon reserves [11], it exerts powerful seasonal radiative forcings. Dark conifer canopies mask reflective snow, lowering surface albedo and generating strong winter–spring warming (+0.5 to +2.5 $W/m^{-2}$) that can offset or surpass the cooling from carbon sequestration [12–14]. Yet boreal forests can also amplify low-level cloud cover (0.1–0.5 %) and emit biogenic volatile organic compounds (BVOCs) that oxidise into secondary organic aerosols (SOA), boosting cloud condensation nuclei (CCN) by 10–100 % [15–17]. This BVOC–SOA–CCN pathway can enhance cloud reflectivity (*the Twomey effect*), producing cooling of –1.8 to –6.7 $W/m^{-2}$ [(17–19)]; a magnitude that could rival, or even outweigh, albedo-driven warming under warmer, cleaner atmospheric conditions. These opposing processes, and the high-stakes uncertainty they create for EEI, are illustrated in the Graphical Abstract (Fig. 1).

The policy stakes are acute: large-scale boreal afforestation deployed without accounting for these counteracting feedbacks risks worsening, rather than mitigating, EEI [20–22]. Yet, this warming–cooling balance remains unresolved because of persistent methodological bottlenecks. Satellite retrievals over snow- and ice-covered landscapes carry large uncertainties [23], Earth System Models (ESMs) struggle to represent BVOC chemistry and mixed-phase cloud processes [19, 24], and observational networks in boreal regions remain sparse and spatially biased [11]. Scale mismatches between plot-based flux towers and coarse-resolution climate models further obscure the non-local atmospheric adjustments that often dominate the net signal [25–26].

This review addresses that gap by providing the first boreal-centric synthesis of afforestation's coupled carbon–albedo–cloud effects on EEI. We integrate satellite observations, flux tower data, paleoclimate reconstructions, and ESM experiments to: (i) disentangle the surface energy, hydrological, and aerosol–cloud pathways through which boreal afforestation alters cloud dynamics; (ii) assess the magnitude, direction, and climatic state-dependence of these feedbacks; and (iii) identify the process-level uncertainties that most limit predictive skill and policy relevance. By reframing boreal afforestation as a biophysically and biogeochemically coupled system, rather than a purely carbon-centric intervention we aim to equip policymakers, land managers, and climate modellers with the evidence base needed to deploy climate-smart forestry that mitigates EEI without unintended radiative penalties.

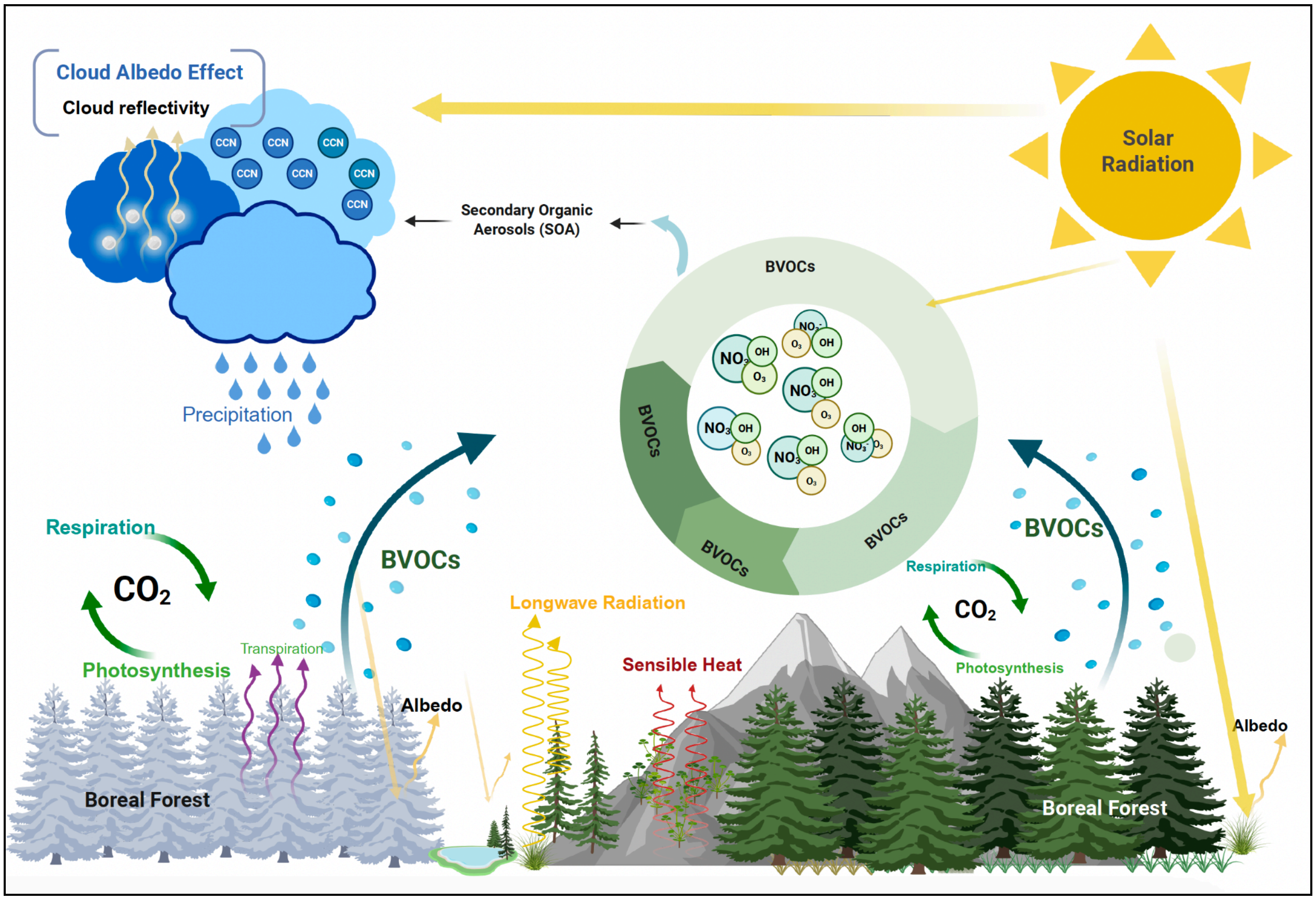

**Fig. 1 Coupled carbon-albedo-cloud pathways of boreal afforestation and Earth's Energy Imbalance.** Schematic of competing biophysical feedbacks by which boreal forests can either warm or cool the climate. Dark canopies mask snow and lower surface albedo, producing positive radiative forcing alongside changes in longwave and sensible-heat fluxes. Conversely, forests emit biogenic volatile organic compounds (BVOCs) that oxidize (OH, $O_3$, $NO_3$) to secondary organic aerosols (SOA), increase cloud-condensation nuclei (CCN), enhance low-level cloud fraction, raise cloud reflectivity (Twomey effect), yielding potential cooling. The BVOC-aerosol-cloud pathway remains poorly constrained yet may strengthen under warmer, cleaner atmospheric conditions, and therefore can rival or exceed albedo-driven warming. These counteracting processes frame the risk that large-scale planting could unintentionally amplify EEI if sited without climate-smart criteria. (Schematic, not to scale. Abbreviations: EEI, Earth's Energy Imbalance; RF, radiative forcing; BVOCs, biogenic volatile organic compounds; SOA, secondary organic aerosols; CCN, cloud-condensation nuclei.). Created in https://BioRender.com.

# Results

## The fundamental biophysical mechanisms connecting tree cover changes and cloud formation.

Forests modulate cloud dynamics through two synergistic pathways: (1) surface energy/water flux modification and (2) biogenic aerosol emissions (Fig. 1, 2).

### Surface Energy and Water Fluxes

Forest canopies enhance evapotranspiration (ET) due to high leaf area indices and deep roots (which allow access to soil moisture reserves unavailable to shallower-rooted plants, particularly during dry periods)[11], elevating atmospheric moisture, which is a prerequisite for cloud formation[12]. The process of converting liquid water to vapor consumes energy (latent heat), leading to a cooling effect at the land surface, particularly during the growing season when ET rates are high[12]. This latent heat flux cools the surface but supplies vapor for cloud condensation[11–12,15]. Concurrently, forest roughness intensifies turbulence[11], promoting vertical heat/moisture transport and convective uplift, particularly for cumulus formation[11,15,27]. While forests exhibit lower albedo (increasing solar absorption and surface warming) [11–12, 105–106], energy partitioning toward ET and sensible heat (H) governs local climate outcomes [11–12]. The Bowen ratio (H/LE) critically influences atmospheric boundary layer (ABL) dynamics and cloud potential [15, 26]. Positive feedbacks may sustain cloud cover through radiation-mediated moisture recycling [15–16]

**Biogenic Aerosol-Cloud Pathway**

Boreal forests emit biogenic volatile organic compounds (BVOCs), predominantly monoterpenes, which oxidize to form secondary organic aerosols (SOA) [13, 44]. SOA particles grow into cloud condensation nuclei (CCN), elevating CCN concentrations by 10 – 100% in boreal regions [12–13, 49]. Via the *Twomey effect*, higher CCN counts yield clouds with more numerous, smaller droplets, enhancing reflectivity (albedo) and exerting net cooling [12, 47]. Though debated, prolonged cloud lifetimes may further amplify cooling [49].

**Observed Patterns, Variability, and Mechanisms**

Satellite observations reveal that afforestation generally enhances low-level cloud cover (LLCC) across 67% of global land areas, with stronger effects during warmer months, suggesting potential surface cooling via increased solar reflection (Fig. 2; Table S1) [15–16]. However, regional variability is significant: LLCC increases over temperate and boreal forests [16, 42, 76] but decreases in the tropics (e.g., Amazon, Central Africa) and parts of the temperate zone (e.g., Southeast US) [16, 42].

This spatial heterogeneity stems from competing biophysical mechanisms. Enhanced sensible heat flux over forests promotes convection and LLCC [15–16, 41]. Conversely, cooler/wetter forests (e.g., vs. adjacent deforested patches) induce mesoscale subsidence, suppressing clouds[16]. Climate models corroborate that large-scale deforestation reduces tropical cloudiness but increases it in boreal/ temperate zones via non-local atmospheric adjustments [58–59]. Forest type also modulates effects; needleleaf forests in Europe enhance LLCC more than broadleaf forests [16].

Palaeoecological evidence underscores forests' active role in climate feedbacks. Holocene northward boreal expansion reduced albedo, amplifying regional warming, while altering moisture cycling and potentially cloud dynamics [114]. Pollen records link tree-line advances to warmer/moister conditions, illustrating persistent vegetation-climate couplings [113–114].

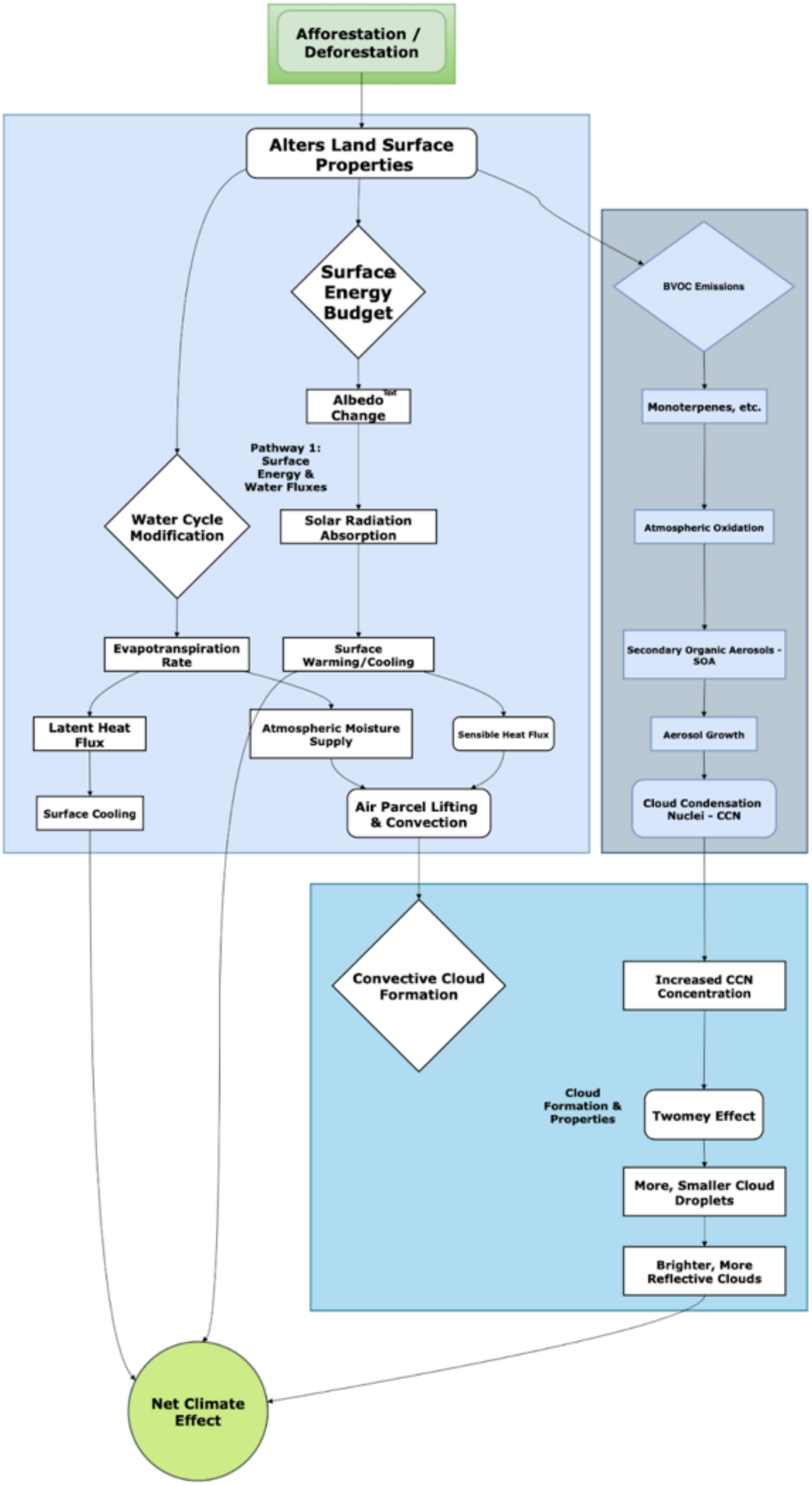

**Fig. 2 | Biophysical mechanisms linking forests to cloud formation.** Forests influence climate through multiple pathways. a, they absorb more solar radiation due to low surface albedo, which provides energy for other processes. b, High evapotranspiration (ET) releases significant water vapor (latent heat flux), moistening the atmosphere. High surface roughness enhances turbulence, mixing heat and moisture into the atmospheric boundary layer (ABL). c, Forests emit biogenic volatile organic compounds (BVOCs). These gases oxidize to form secondary organic aerosol (SOA), which grow into cloud condensation nuclei (CCN). d, the increased abundance of CCN and water vapor promotes the formation of clouds with more, smaller droplets, which are more reflective (higher albedo), exerting a cooling effect.

## Impacts of boreal afforestation on local and regional cloud dynamics, surface albedo, and energy fluxes

### Local and Non-local Atmospheric Effects

Forest cover changes trigger climatic responses at local and non-local scales via altered surface properties (albedo, roughness, evapotranspiration [ET]) and energy fluxes. Local effects modify near-surface temperature, humidity, and turbulence, directly influencing cloud formation/dissipation within the affected area [27–30]. Non-local effects, driven by large-scale atmospheric circulation shifts (e.g., teleconnections), alter cloud cover and precipitation remotely [26, 29, 58–59]. For instance, deforestation can perturb weather patterns downstream, such as displacing the Intertropical Convergence Zone [58–59]. CMIP6 models indicate non-local cloud enhancement in boreal/temperate zones post-deforestation due to cooling and moistening, contrasting with local cloud reduction in the tropics [58–59]. Discrepancies between observations (local cloud inhibition [16, 42]) and models (non-local dominance) [58–59] underscore scale- and methodology-dependent interpretations of forest-cloud impacts.

### Boreal Albedo Effect: Snow Masking and Seasonality

Boreal afforestation reduces surface albedo via snow masking, where dark conifer canopies obscure reflective snow, boosting shortwave absorption (+14 – 23% annually in Finnish forests vs. peatlands )[35]. This effect peaks in winter/spring under snow cover but weakens in summer due to higher ET-driven cooling [12, 35]. Albedo varies with forest structure (disturbances (e.g., fire)) or deciduous stands increase albedo, while conifers exacerbate warming [36,107]. Climate change shortens snow seasons, amplifying albedo-driven warming via positive feedback and treeline expansion [36,105] (Fig. 1).

### Cloud Dynamics and BVOC-Aerosol Pathway

Boreal forests enhance cloud formation through both biophysical and biochemical pathways. Observational evidence shows that air masses over Fennoscandia exhibit increased cloud condensation nuclei (CCN) concentrations by over 100 $cm^{-3}$, along with heightened humidity and cloud optical thickness after more than 50 hours of exposure to forested landscapes, a response linked to sensible heat fluxes[16,49]. Modeling results from CMIP6 simulations further indicate non-local increases in cloud cover following deforestation[61], while afforestation produces regionally variable cooling or warming effects due to the trade-off between surface albedo and evapotranspiration[11]. Additionally, biogenic volatile organic compounds (BVOCs), particularly temperature-sensitive monoterpenes[47], contribute to cloud formation by generating secondary organic aerosols (SOAs), which can double CCN concentrations[13]. This process increases cloud droplet numbers, enhancing cloud reflectivity via the *Twomey effect* and producing a cooling influence ranging from –1.8 to –6.7 $W/m^2$ [13]. Remote sensing data support this mechanism, showing aerosol-induced cloud brightening [47].

### Net Energy Balance and State Dependency

Boreal afforestation's net climate impact is shaped by competing feedbacks. On one hand, it induces warming through reduced surface albedo, particularly in winter and spring due to snow masking effects [35]. On the other hand, it promotes cooling through enhanced summer evapotranspiration and aerosol-mediated cloud brightening [12–13].

Satellite observations suggest that, on an annual basis, boreal forests contribute to net warming, as strong cold-season albedo effects outweigh the cooling observed in summer [30]. However, model simulations reveal substantial regional variability in this outcome [11]. Notably, the biogenic volatile organic compound (BVOC)-aerosol pathway could shift the balance toward net cooling under warmer conditions, when BVOC emissions are at their peak, thereby potentially countering albedo-induced warming [13, 47] (Fig. 2).

While afforestation in boreal zones reduces surface albedo through snow masking, it simultaneously enhances cloud formation locally via aerosol emissions and non-locally through changes in atmospheric circulation. The overall impact on Earth's energy balance remains uncertain, as albedo-induced warming may be offset by BVOC-driven cloud cooling in a warming climate. The resolution of this uncertainty depends on improved integration of scale-dependent methodologies (e.g., observations versus models) and greater understanding of state-dependent feedback mechanisms.

## Afforestation's Net Climate Forcing via Carbon and Cloud Pathways

### Net Climate Impact of Boreal Afforestation on EEI

Evaluating boreal afforestation's net effect on Earth's Energy Imbalance (EEI) hinges on quantifying radiative forcing (RF) from three competing processes: carbon sequestration (cooling), surface albedo changes (warming), and cloud alterations (variable cooling) [65].

### The Core Trade-off: Albedo Warming vs. Cloud Cooling

Albedo-Driven Warming: Boreal afforestation replaces bright, snow-covered landscapes with dark canopies, inducing strong positive RF via "snow masking." Observational and modeling studies estimate +2.3 ± 2.2 W/m² for mature forests relative to fire-cycled landscapes (Table 3) [13, 35]. This spatially heterogeneous warming remains a critical policy challenge [65].

Cloud-Aerosol Cooling: BVOC emissions enhance CCN concentrations (Fig. 3c), increasing cloud reflectivity (*Twomey effect*). Modeling suggests substantial local negative RF (–1.8 to –6.7 W/m²; Table 3), potentially dominating albedo warming [13]. This feedback is temperature-dependent: warming amplifies BVOC emissions, strengthening cooling (–0.43 W/m² globally in feedback-on models; Table 3) [13, 54].

### Latitudinal Contrasts and Net RF Uncertainty

Biome-specific outcomes reveal contrasting climate effects across forest types (Fig. 3e). In boreal and temperate coniferous regions, afforestation tends to result in net warming, as the albedo-driven warming outweighs the cooling effects from carbon uptake and cloud-related processes[14,112]. In contrast, tropical forests exhibit robust net cooling, where high carbon sequestration combined with strong evapotranspiration and cloud feedbacks more than compensate for the loss of surface albedo[11]. The boreal context remains particularly contentious. Simulations of deforestation in these regions suggest a net cooling effect driven by albedo gains, amounting to approximately –0.8 K by 2100 [29] (see Table 3). However, models that incorporate BVOC emissions suggest that under warming conditions, these emissions could lead to a shift toward net cooling[13]. Divergences in how models represent cloud and albedo processes contribute significantly to the uncertainty in projections, as illustrated by the wide radiative forcing spread in CMIP6 simulations (ranging from 0 to –0.17 $W/m^2$) [61, 76].

## EEI Stabilization Risks and Urgency

The climate implications of boreal afforestation are critical and hinge on the balance between competing radiative forces. If afforestation results in net warming, it could exacerbate Earth's Energy Imbalance (EEI), thereby negating the intended carbon sequestration benefits[13],[19]. Conversely, if net cooling prevails, it could help offset regional contributions to the global decline in planetary albedo[6]. However, this potential benefit is undermined by unresolved uncertainties surrounding cloud feedbacks, particularly those mediated by biogenic aerosols[47].

From a policy standpoint, temperature-dependent BVOC-cloud feedbacks are likely to intensify in cleaner atmospheric conditions[57]. Nevertheless, large-scale boreal afforestation remains a questionable strategy in the absence of greater clarity regarding radiative forcing dynamics. Ultimately, the net impact of boreal afforestation on EEI, driven by the interplay between albedo-induced warming and cloud-aerosol-driven cooling represents a high-stakes uncertainty that demands urgent scientific resolution.

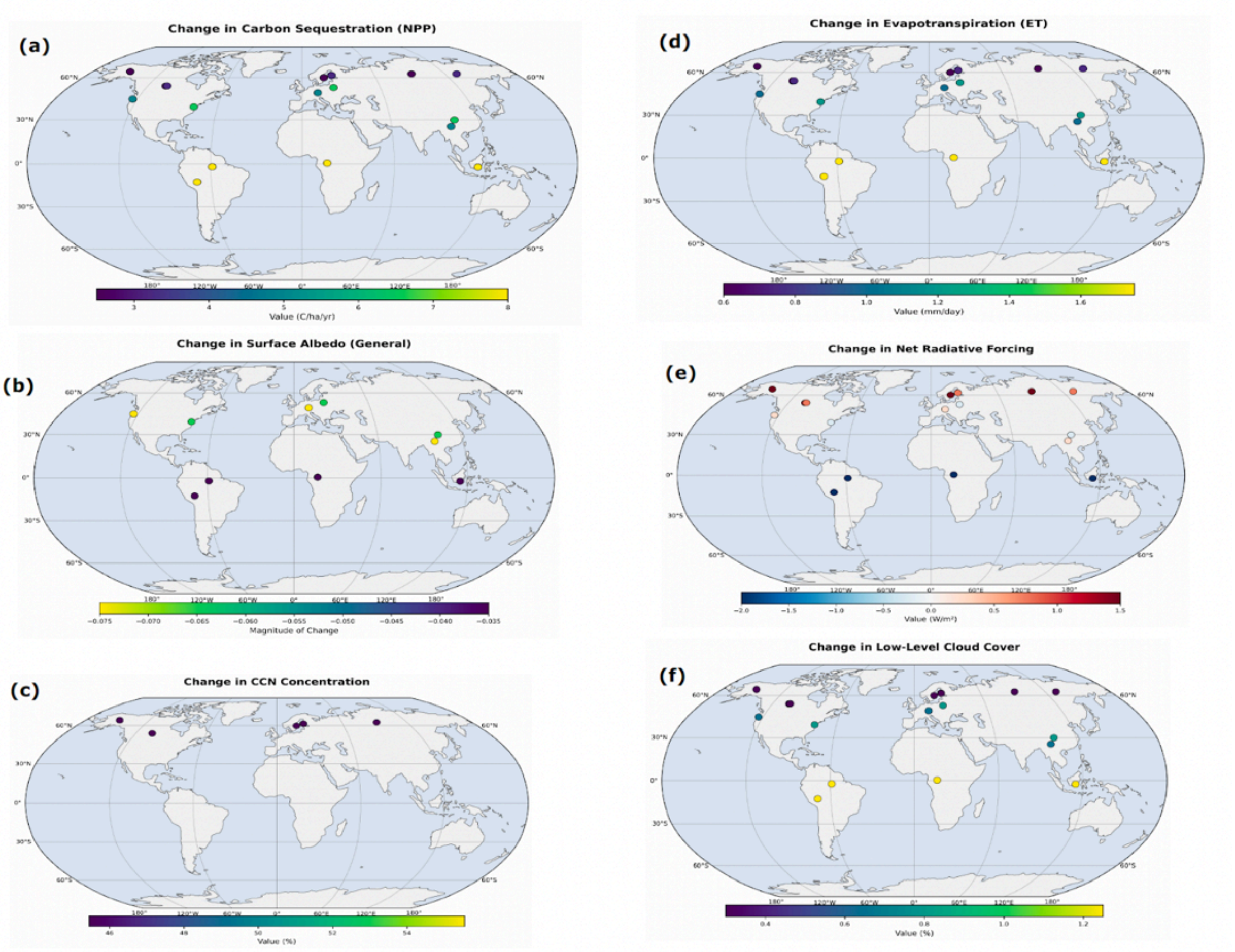

**Fig. 3 | Biophysical and biogeochemical impacts of afforestation.** The maps illustrate the estimated annual changes in key biophysical and biogeochemical variables resulting from afforestation. The data points correspond to the locations of empirical and observational studies, including flux tower networks, which form the evidence base for this review. The panels show changes

in: (a) Carbon Sequestration as Net Primary Productivity (NPP), measured in carbon per hectare per year (C/ha/yr). (b) Surface Albedo, showing the magnitude of change. (c) Cloud Condensation Nuclei (CCN) concentration, as a percentage change. (d) Evapotranspiration (ET), in millimeters per day (mm/day). (e) Net Radiative Forcing, in watts per square meter (W/m²). (f) Low-Level Cloud Cover, as a percentage change. Each panel includes a colorimetric scale indicating the magnitude and direction of the change for the respective variable.

**Table 1 | Comparison of Estimated Radiative Forcing (RF) Components for Boreal Afforestation and Deforestation.** Summarizes key studies quantifying RF from carbon sequestration, surface albedo changes, and cloud feedback. Negative RF indicates cooling; positive RF indicates warming. Values are study-specific and not directly comparable due to methodological differences (e.g., scale, land-cover transition).

| Study/ Model Type | Region/Scale | Forcing Component | Estimated RF (W/m²) / Effect | Key Assumptions /Notes | Source (s) |
|---|---|---|---|---|---|
| Global C-Cycle /Climate Model (INCCA) | Boreal Deforestation | Net (Carbon + Biophysical) | Net Global Cooling (~0.8 K by 2100 vs Standard) | Complete boreal deforestation; albedo increase dominates $CO_2$ release warming. | 30 |
| Global C-Cycle /Climate Model (INCCA) | Global Deforestation | Net (Carbon + Biophysical) | Net Global Cooling (~0.3 K by 2100 vs Standard) | Strongest cooling from boreal removal. | 30 |
| Global Model / Synthesis | Boreal Forestation | Surface Albedo Change | Positive RF (Warming) | Offsets carbon sequestration benefit partially or fully. | 14, 105-106 |
| Observation/ Model Analysis (Randerson+) | Boreal Forest | Surface Albedo (Mature vs Fire Cycle) | +2.3 ± 2.2 W/m² (Warming) | Compares mature forest to 80-year fire cycle landscape. | 13 |
| Global Aerosol Model (GLOMAP) / Rad. Model | Boreal Forest | Cloud Albedo (BVOC-Aerosol Effect) | -1.8 to -6.7 W/m² (Cooling) | Based on doubling of CCN (100->200 $cm^{-3}$); potentially offsets albedo warming. | 13 |
| ESM (NorESM) | Global / Future | Net Cloud Forcing (BVOC Feedback) | -0.43 W/m² (Cooling enhancement) | Compares feedback-on vs. feedback-off in $2xCO_2$ scenario; stronger effect with low anthropogenic aerosols (-0.73 W/m²). | 55 |
| Satellite LST Analysis | Boreal Forests | Net Biophysical Effect (LST) | Net Warming Annually | Strong winter warming (albedo) outweighs moderate summer cooling (ET). | 31 |
| CMIP5 Models | Historical LCC | Albedo-Induced RF | -0.07 W/m² (Global Mean Cooling) (Range 0 to -0.17) | Based on reconstructed historical deforestation (mainly mid-latitude/ tropical); model biases identified. | 71 |
| Integrated Assessment Model (GCAM/CESM) | Future LULCC | Albedo-Induced RF | -0.06 to -0.29 W/m² (Global, by 2070) | pends on scenario (afforestat vs. bioenergy); highlights policy relevance of albedo forcing. | 67 |

*Note: RF values are often specific to the study's methodology, scale, and assumptions (e.g., local vs. global, specific land cover transition). Negative RF indicates cooling, positive RF indicates warming.*

## Identifying critical knowledge gaps, modeling challenges, and observational constraints hindering a complete understanding, especially concerning boreal forest-cloud interactions.

### Knowledge Gaps and Challenges

Boreal forest–cloud interactions remain poorly constrained by both models and observations. Earth System Models (ESMs) exhibit several critical deficiencies: cloud parameterizations struggle to accurately represent low-level and mixed-phase clouds [6,46,72]; the BVOC–SOA–CCN pathway is marked by considerable uncertainty in emission factors, secondary organic aerosol (SOA) yields, and cloud condensation nuclei (CCN) activation [42,45,5 0]; persistent surface albedo biases, especially those related to snow masking, arise from inadequate representations of vegetation structure [35, 105]; and land–atmosphere coupling fails to capture important non-local teleconnections[14, 58]. These limitations collectively contribute to a wide spread in simulated cloud radiative forcing and lead to divergent estimates of the net climate forcing from boreal afforestation[11, 13, 26] (Table 3).

Observational constraints present additional challenges. Satellite retrievals of aerosol and cloud properties over snow- and ice-covered regions are indirect and prone to substantial errors [56,105] while ground-based measurements such as flux towers remain sparse and spatially limited in remote boreal landscapes[48,56] (Figs. 3, S1). Key process-level gaps persist in several areas: BVOC and SOA chemistry, particularly concerning oxidation pathways and particle hygroscopicity under boreal conditions [48]; aerosol activation and the efficiency of CCN formation in cold, clean air masses [56,70]; the influence of forest species composition and stand age on albedo and energy partitioning[17,33] ; the long-term effects of disturbances such as fire and insect outbreaks on cloud–albedo feedbacks[20,33]; and the interactions between permafrost dynamics and vegetation structure, especially with respect to soil carbon stability under afforestation scenarios [105].

### Indigenous Knowledge as a Critical Missing Piece

Indigenous knowledge systems, which encompass millennia of place-based ecological observations, remain largely excluded from mainstream scientific research[95], resulting in a significant gap in understanding complex, non-linear climate feedbacks. This exclusion limits insights into several critical processes, such as the historical recurrence of fire and insect disturbances and their effects on local microclimates[20,95]; the species-specific biophysical influences of coniferous versus deciduous canopies on snow albedo and moisture recycling[36,95]; and the interactions between vegetation shifts and permafrost stability, particularly regarding ground-ice dynamics[17,95].

Neglecting these knowledge systems impedes the contextually grounded validation of Earth system models and satellite retrievals, as emphasized in Table 4.

### Implications for Earth's Energy Imbalance

Collectively, these gaps, including the underutilization of Indigenous knowledge impede robust quantification of boreal afforestation's RF and its influence on EEI [6, 13, 70]. The exclusion of holistic, long-term ecological insights perpetuates uncertainties in climate mitigation strategies [22, 95].

**Table 2 | Summary of Key Uncertainties and Knowledge Gaps Hindering Assessment of Boreal Forest-Climate Feedback.** Identifies critical limitations in modeling, observations, and process understanding that contribute to uncertainty in quantifying the net climate impact of boreal afforestation, particularly regarding cloud dynamics, albedo, and permafrost interactions. Includes implications for evaluating radiative forcing and EEI stabilization.

| Area of Uncertainty | Specific Gap/Challenge | Implication for Net Climate Effect Assessment | Source (s) |
|---|---|---|---|
| Cloud Parameterization | Poor representation of low-level and mixed-phase clouds; sub-grid scale convection; aerosol activation physics. | High uncertainty in magnitude and sign of cloud radiative forcing and feedbacks associated with afforestation. | [6] |
| BVOC/SOA Processes | Uncertainties in BVOC emission rates (especially under stress), SOA chemical pathways, yields, and hygroscopic properties. | Difficulty in modelling SOA formation, CCN concentrations, and the strength of the aerosol–cloud albedo effect. | [42] |
| Aerosol–Cloud Interactions | Challenges in quantifying Nd susceptibility to CCN in clean/cold environments; difficulty separating aerosol effects from meteorology. | Large uncertainty in magnitude of cooling from biogenic aerosol-enhanced cloud reflectivity. | [45] |
| Surface Albedo Modelling | Inaccurate representation of snow–vegetation interactions, canopy structure (LAI/PAI), forest type, and topographic effects. | Biases in estimating positive radiative forcing (warming) from reduced surface albedo after afforestation. | [17] |
| Land Surface Models (LSMs) | Uncertainties in evapotranspiration, roughness, soil moisture, and phenology parameterizations. | Affects simulation of surface energy/water balance, influencing atmospheric conditions for cloud formation and local temperature responses. | [18] |
| Observational Constraints | Limited spatial coverage of ground data; satellite retrieval limitations (vertical profiles, CCN proxies, snow detection). | Difficulty validating models and constraining key processes at relevant scales, especially for aerosol–cloud interactions. | [42] |
| Non-local Effects & Coupling | Poorly constrained large-scale atmospheric responses (teleconnections) to regional land-cover change. | Uncertainty in regional and global climate impacts beyond immediate afforestation areas; potential for unexpected remote effects. | [58] |
| Permafrost Interactions | Complex feedbacks between afforestation, ground thermal regime, thaw dynamics, hydrology, and carbon release ($CO_2$, $CH_4$). | High uncertainty in long-term carbon balance and net climate effect; risk of triggering large carbon emissions. | [17] |
| Disturbance Regimes | Increasing fire and insect impacts under climate change; effects on succession, carbon stocks, albedo, and feedbacks. | Uncertainty in long-term stability of carbon storage and biophysical properties of afforested boreal landscapes. | [20] |
| **Indigenous Knowledge Integration** | Limited incorporation of Indigenous ecological knowledge and long-term, place-based observations into modelling frameworks. | Reduces contextual accuracy, omits historical disturbance and land-use patterns, and limits ability to validate non-linear feedbacks such as snow–albedo–vegetation interactions and permafrost stability. | [95, 20, 17] |

# Discussion

## Policy and Management Implications for Climate Stabilization

Our synthesis underscores that boreal afforestation presents a complex trade-off for climate stabilization, with competing radiative forcings that vary in magnitude, sign, and spatial distribution. The results (Figs. 2–3; Table 3) confirm strong albedo-driven warming (+0.5 to +2.5 W $m^{-2}$) from snow masking in coniferous stands [13, 35], offset in part by BVOC-cloud cooling (–1.8 to –6.7 W $m^{-2}$) mediated by secondary organic aerosols and enhanced cloud condensation nuclei [13]. Yet, the net radiative effect remains indeterminate, with current Earth System Models (ESMs) unable to fully resolve BVOC–SOA–CCN chemistry, mixed-phase cloud dynamics, and teleconnection-driven non-local effects [42, 45, 58]

This uncertainty is not merely academic; it carries tangible policy risks. Large-scale initiatives, such as Canada's *2 Billion Tree Commitment* [19], could inadvertently worsen Earth's Energy Imbalance (EEI) if afforestation in high-snow, carbon-rich peat or permafrost regions amplifies albedo warming [35, 105]. Climate-positive deployment demands spatially explicit siting on degraded or low-snow landscapes, prioritization of mixed-species plantings to increase winter albedo and diversify BVOC emission profiles [36, 83, 107], and explicit accounting of biophysical forcing in carbon markets [28, 65].

The temperature dependence of BVOC emissions (5–20 % $°C^{-1}$ increase; [13]) further complicates this balance. Under warmer, cleaner atmospheric conditions, these emissions may strengthen cloud brightening [54], but empirical constraints remain sparse. Until these processes are reliably parameterized in models and validated against multi-scale observations, boreal afforestation cannot be treated as a universally climate-positive intervention.

### Proposed Framework for Assessing Net Climate Impact of Boreal Afforestation

Figure 4 presents a tiered decision-support framework that integrates biogeochemical (carbon sequestration, non-$CO_2$ GHG fluxes, disturbance risks) and biophysical processes (albedo, evapotranspiration, BVOC–SOA–CCN–cloud pathways, non-local circulation effects). The framework operationalizes the calculation of net radiative forcing:

$$\Delta EEI = RF_C + RF_A + RF_{Cloud}$$

where $RF_C$, $RF_A$, and $RF_{Cloud}$ represent carbon, albedo, and cloud-mediated forcing components respectively.

In synthesizing the proposed framework, several assumptions guided the integration of these processes. Disturbance regimes such as fire recurrence and insect outbreaks, and permafrost dynamics were represented using averaged historical patterns, despite their inherently state-dependent nature[17, 20]. Estimates of BVOC emission factors and SOA yields were drawn from established field programmes, including SMEAR and BOREAS, but may underestimate potential responses under extreme warming or drought [13, 42]. Albedo effects were evaluated by contrasting mature forest stands with fire-cycled landscapes to capture structural and phenological variability [13, 35]. Non-local cloud responses were incorporated from CMIP6-LUMIP ensemble means, although the large inter-model spread underscores persistent uncertainties in simulating atmospheric teleconnections [58, 71]. These assumptions, while necessary to operationalize the framework, point directly to the need for integrated modelling tools such as CBM-CFS3 for carbon accounting [116], GLOMAP and NorESM[117] for aerosol–cloud interactions, and CLASS-CTEM[118] for surface–permafrost dynamics that can couple these processes and reduce uncertainty in estimating net radiative forcing.

The framework also recognises that forests deliver co-benefits aligned with multiple Sustainable Development Goals (SDGs) including biodiversity conservation (SDG 15), economic prosperity (SDG 8), food and water security (SDGs 2 and 6), and renewable energy potential (SDG 7) [119–121]. Similarly, afforestation goals such as timber production, bioenergy, or rural livelihoods, though outside the primary climate-centric scope of this review, must be integrated into holistic land-use planning.

**Climate-Smart Forestry Imperatives**

Delivering climate mitigation without triggering ecological maladaptation in the boreal zone depends on a set of complementary operational principles. Afforestation efforts should avoid high-risk sites, including peatlands, permafrost, and long-duration snow landscapes, where biophysical feedbacks tend to amplify warming [35, 110]. Equally important is the diversification of species and stand structure, with an emphasis on incorporating deciduous species to increase snow-season albedo and enhance seasonal resilience [36, 112]. Process-based metrics that extend beyond carbon sequestration such as changes in surface albedo, evapotranspiration partitioning, and cloud condensation nuclei enhancement must be integrated into Monitoring, Reporting, and Verification (MRV) systems. Together, these measures align with the principles of "nature-based solutions" [22, 93], ensuring that boreal afforestation supports the stabilization of Earth's Energy Imbalance while safeguarding biodiversity and cultural values.

**Indigenous Knowledge and Research Frontiers**

The exclusion of Indigenous knowledge systems [93] from boreal climate modelling limits the ability to capture long-term, non-linear feedbacks essential for accurate ΔEEI assessment. Indigenous observations offer critical insights into disturbance histories [20, 93], species-specific snow–albedo dynamics [36, 93], and vegetation–permafrost interactions [17, 93] that remain poorly represented in Earth System Models. Embedding Indigenous-led monitoring within the proposed framework (Fig. 4) strengthens model validation, improves contextual accuracy, and ensures equitable co-benefit distribution. Closing current gaps demands targeted process studies on BVOC–SOA–CCN interactions under boreal thermal and radiation regimes [44, 48, 75], tighter coupling of permafrost and hydrology models with forest growth and disturbance modules, and integrated flux tower–satellite campaigns across disturbance gradients [48, 52] to bridge model–observation scale mismatches. Developing coupled decision-support tools that explicitly link climate mitigation potential to Sustainable Development Goal co-benefits will be essential for ensuring boreal afforestation strategies are scientifically credible and socially inclusive.

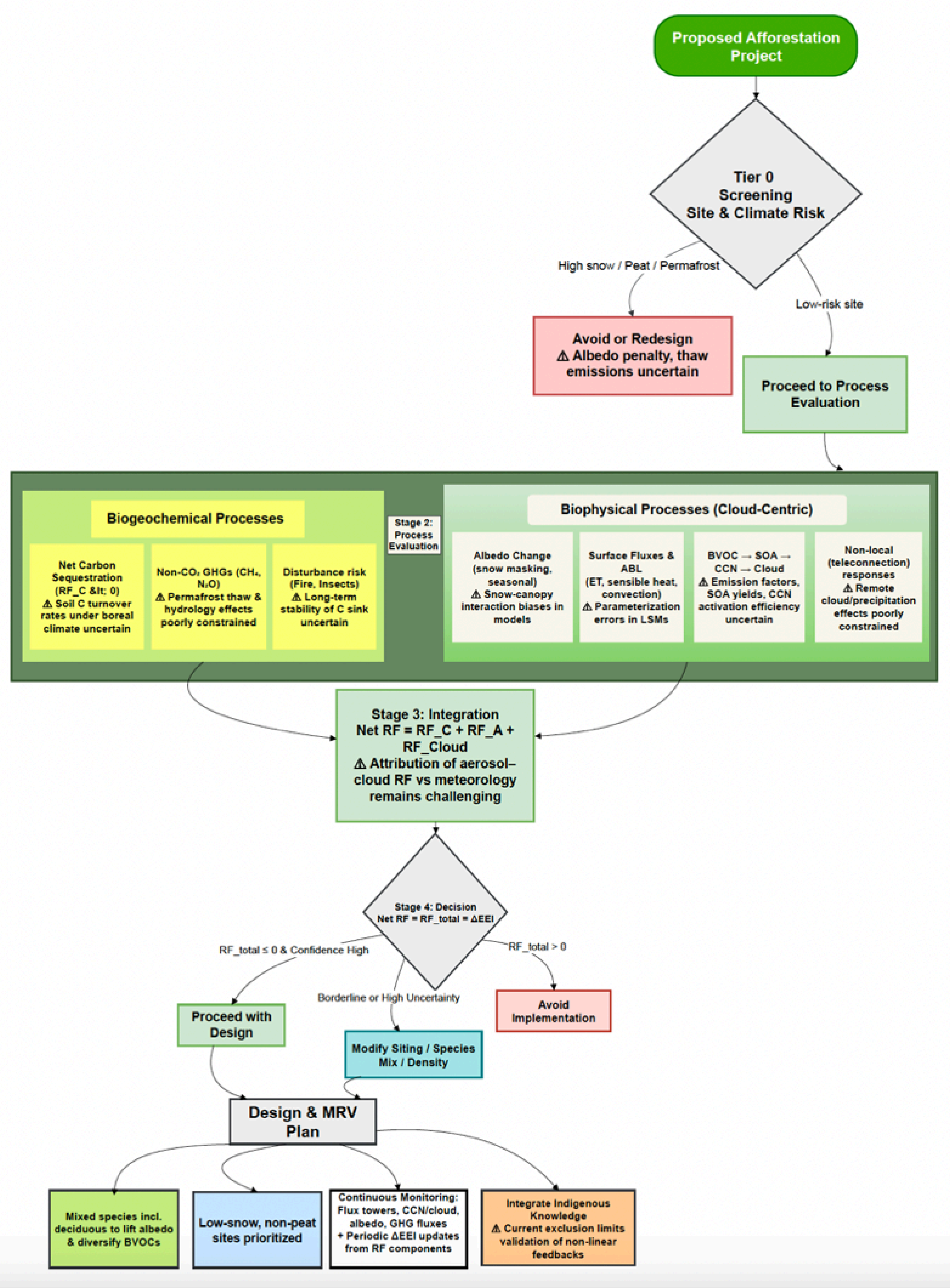

Proposed Afforestation Project
Tier 0 Screening Site & Climate Risk
High snow / Peat / Permafrost
Low-risk site
Avoid or Redesign ⚠ Albedo penalty, thaw emissions uncertain
Proceed to Process Evaluation
Biogeochemical Processes
Stage 2: Process Evaluation
Biophysical Processes (Cloud-Centric)
Net Carbon Sequestration (RF_C < 0) ⚠ Soil C turnover rates under boreal climate uncertain
Non-$CO_2$ GHGs ($CH_4$, $N_2O$) ⚠ Permafrost thaw & hydrology effects poorly constrained
Disturbance risk (Fire, Insects) ⚠ Long-term stability of C sink uncertain
Albedo Change (snow masking, seasonal) ⚠ Snow-canopy interaction biases in models
Surface Fluxes & ABL (ET, sensible heat, convection) ⚠ Parameterization errors in LSMs
BVOC → SOA → CCN → Cloud ⚠ Emission factors, SOA yields, CCN activation efficiency uncertain
Non-local (teleconnection) responses ⚠ Remote cloud/precipitation effects poorly constrained
Stage 3: Integration Net RF = RF_C + RF_A + RF_Cloud ⚠ Attribution of aerosol–cloud RF vs meteorology remains challenging
Stage 4: Decision Net RF = RF_total = ΔEEI
RF_total ≤ 0 & Confidence High
Borderline or High Uncertainty
RF_total > 0
Proceed with Design
Modify Siting / Species Mix / Density
Avoid Implementation
Design & MRV Plan
Mixed species incl. deciduous to lift albedo & diversify BVOCs
Low-snow, non-peat sites prioritized
Continuous Monitoring: Flux towers, CCN/cloud, albedo, GHG fluxes + Periodic ΔEEI updates from RF components
Integrate Indigenous Knowledge ⚠ Current exclusion limits validation of non-linear feedbacks

**Figure 4: A proposed framework for assessing the net climate impact of afforestation in the Boreal Ecosystem.** The process begins with a risk screening for factors like permafrost and snow. It then evaluates competing biogeochemical (e.g., carbon sequestration) and biophysical (e.g., albedo, cloud formation) effects. An integrated assessment of the total radiative forcing determines whether a project should proceed, be modified, or avoided. The final stage outlines a robust design and monitoring plan to ensure intended climate outcomes. Within the framework, the warning symbol **(⚠)** denotes uncertainty or a change in a given factor. RF stands for Radiative Forcing (with subscripts C, A, and Cloud for Carbon, Albedo, and Cloud components), which is equivalent to the change in Earth's Energy Imbalance (ΔEEI). Other terms include the greenhouse gases methane ($CH_4$) and nitrous oxide ($N_2O$); the aerosol pathway components Biogenic Volatile Organic Compounds (BVOC), Secondary Organic Aerosols (SOA), and Cloud Condensation Nuclei (CCN); Land Surface Models (LSMs); and MRV for Monitoring, Reporting, and Verification.

# Conclusion

Boreal afforestation remains a conditional and context-dependent climate strategy. While its carbon sequestration potential is significant, our synthesis shows that albedo-driven warming, especially from snow masking, can offset or reverse these gains unless counterbalanced by robust cloud-mediated cooling via BVOC–SOA–CCN pathways. This cooling mechanism, though potentially decisive under warmer and cleaner atmospheric conditions, remains one of the least constrained elements in boreal climate science.

Given the accelerating Earth's Energy Imbalance, partly linked to reduced planetary albedo from cloud decline, indiscriminate deployment of boreal afforestation risks exacerbating rather than mitigating warming. Effective action requires a climate-smart approach: prioritizing degraded, low-snow sites; diversifying species to improve seasonal albedo and resilience; embedding biophysical metrics into MRV systems; and integrating Indigenous knowledge to capture long-term, non-linear feedbacks.

Until BVOC-cloud feedbacks are empirically constrained and reliably represented in Earth System Models, boreal afforestation should not be positioned as a primary solution for EEI stabilization. Instead, it should be implemented selectively within an integrated land-use strategy that balances carbon, albedo, and cloud effects; so that it functions as a credible, verifiable component of holistic climate stewardship rather than a speculative geoengineering measure.

## Supplementary

Table S1 presents a synthesis of the competing climatic fingerprints of global forest transitions across six major biomes, quantifying how afforestation or deforestation influences Earth's energy balance through interconnected biophysical and biogeochemical pathways. The dataset integrates peer-reviewed modelling studies, flux tower observations, and satellite retrievals conducted between 2000 and 2024 [11, 13, 14, 15, 16, 26, 42, 47, 76, 106, 107, 108], spanning ecosystems from boreal to tropical zones. It cross-compares ten key climate drivers, ranging from seasonal albedo shifts to BVOC-mediated cloud effects, using biome-specific response ranges. Reported ranges incorporate uncertainty arising from variability in species composition, microclimate, and land-use history, with net radiative forcing values sometimes differing by as much as ±0.5–2.0 $W/m^2$ [(14, 26, 106)].

In boreal coniferous forests, dark evergreen canopies mask underlying snow during winter, creating a pronounced albedo penalty that drives strong warming, equivalent to a +0.3–0.6 increase in albedo with deforestation [14, 26]. Although these forests emit high levels of BVOCs that can elevate CCN concentrations by 10 –100% [13, 47], the resulting cloud-related cooling is generally outweighed by snow masking, producing a net positive forcing of +0.5 to +2.5 $W/m^2$ [11, 14]. In contrast, tropical evergreen forests consistently exert a cooling influence, with high evapotranspiration rates of 3–5 mm/day and enhanced low-level cloud cover (±0.5–2.0%) driving net negative forcing of −1.0 to −3.0 $W/m^2$ [15, 16, 42]. Deforestation in these regions releases substantial amounts of carbon; 100 to over 250 tC/ha, representing the highest carbon emissions among the biomes examined [107]. Temperate deciduous forests exhibit a more balanced seasonal pattern; moderate summer albedo values (0.12–0.20) partially offset winter gains, producing near-neutral annual radiative forcing between −1.5 and 0 $W/m^2$ [(26)].

Comparisons across biomes reveal clear contrasts: boreal forests tend toward net warming of approximately +2.5 $W/m^2$, while tropical forests achieve cooling of about −3.0 $W/m^2$ [[14, 26]]. Low-level cloud cover responses are also divergent, with boreal regions showing only minor changes (±0.5%) and tropical zones demonstrating strong amplification (±2.0%) [16, 42, 76, 106]. The carbon–biophysics trade-off is particularly stark; in boreal systems, carbon sequestration gains of 1– 4 tC/ha/yr are often overwhelmed by albedo-driven warming [107], whereas tropical forests achieve synergistic cooling through combined carbon uptake and biophysical feedback [15, 16, 42].

From a policy perspective, this synthesis underscores the need for biome-specific strategies. Boreal afforestation carries a significant risk of net warming despite its carbon uptake potential (NPP: 1– 4 tC/ha/yr), making climate-smart site selection essential [107]. In contrast, tropical afforestation offers maximum co-benefits by combining high carbon storage rates (4 – 12 tC/ha/yr) with robust biophysical cooling [15, 16, 42]. Temperate deciduous and mixed stands provide a middle ground, delivering balanced radiative forcing outcomes and resilience in long-term carbon sinks [26]. These contrasts are spatially contextualised and visually summarised in Fig. 3 of the main text, linking quantitative biome responses to geospatial patterns in the global forest–climate system.

**Table S1 | Biophysical and biogeochemical impacts of afforestation and deforestation on key climate variables across diverse forest biomes.** This table presents a quantitative summary of the effects of different forest types on surface albedo for both winter and summer, evapotranspiration (ET), sensible heat flux, low-level cloud cover (LLCC), biogenic volatile organic compound (BVOC) emissions, and cloud condensation nuclei (CCN) concentrations. Furthermore, it includes data on carbon sequestration rates (Net Primary Productivity, NPP) and potential carbon emissions from above-ground (AG) biomass due to deforestation. The net biophysical radiative forcing, which integrates the climatic effects of albedo and energy fluxes, is also presented. The data are compiled from a range of modeling studies and empirical observations as cited.

| Metric | Boreal Coniferous | Boreal Deciduous | Temperate Coniferous | Temperate Deciduous | Temperate Mixed | Tropical Evergreen | References & Notes |
|---|---|---|---|---|---|---|---|
| Surface Albedo (Winter) | 0.10 to 0.50 → +0.3 to +0.6 with deforestation | 0.20 to 0.50 → +0.1 to +0.3 | 0.08 to 0.15 → ±0.05 to 0.10 | 0.15 to 0.25 → slight changes | 0.10 to 0.18 → ±0.04 to 0.09 | 0.11 to 0.14 → ±0.02 to 0.05 | [14, 26] |
| Surface Albedo (Summer) | 0.05 to 0.12 → ±0.05 - 0.10 | 0.12 to 0.18 → ±0.03 - 0.08 | N/A | 0.12 to 0.20 → ±0.03 to 0.08 | N/A | N/A | [26] *General albedo values* |
| Annual Avg. Albedo Change | ±0.02 to 0.1 | N/A | N/A | N/A | N/A | N/A | [11] |
| Evapotranspiration (ET) | 1.0 to 3.0 mm/day → ±0.2 to 1 | 1.5 to 3.5 mm/day → ±0.3 to 1.2 | 2 to 4 mm/day → ±0.5 to 1.5 | 3 to 6 mm/day → ±1 to 2 | 2.5 to 4.5 mm/day → ±0.7 to 1.7 | 3 to 5 mm/day → ±1 to 2.5 | [15] |
| Sensible Heat Flux Change | ↓ with afforestation, ↑ with deforestation | N/A | ↓ with afforestation, ↑ with deforestation | N/A | N/A | ↓ with afforestation, ↑ with deforestation | [26] |
| Low-Level Cloud Cover (LLCC) Change | ±0.1 to 0.5% | ±0.1 to 0.5% | ±0.3 to 1.0% | ±0.5 to 1.5% | ±0.4 to 1.2% | ±0.5 to 2.0% | [15, 16, 71, 101] |
| BVOC Emissions | High → Increase/Decrease | Variable → Increase/Decrease | High → Increase/Decrease | High → Increase/Decrease | N/A | Very High → Strong Increase/Decrease | [13] *Species-dependent* |
| CCN Concentration Change | +10% to +100% | N/A | N/A | N/A | N/A | N/A | [13] |
| Carbon Sequestration (NPP) | 1.0 to 4.0 tC/ha/yr | 1.5 to 5.0 tC/ha/yr | 2.0 to 8.0 tC/ha/yr | 3.0 to 10.0 tC/ha/yr | 3.0 to 10.0 tC/ha/yr | 4.0 to 12.0 tC/ha/yr | [102, 103] |
| Carbon Emissions (Deforestation) | 50 to 150 tC/ha (AG) + soil | N/A | N/A | N/A | N/A | 100 - 250+ tC/ha (AG) | [102] |
| Net Radiative Forcing (Biophysical) | +0.5 to +2.5 W/m² | 0 to ±1.5 W/m² | -0.5 to +1.0 W/m² | -1.5 to 0 W/m² | -1.0 to +0.5 W/m² | -1.0 to -3.0 W/m² | [14, 11, 26, 101] |

*Note: The values are approximate and context-dependent, derived from regional and global literature as referenced. Where applicable, ranges reflect variability across species, seasonality, and prior land cover. Symbols used in the table are defined as follows: →: Change from baseline; ±: Represents a range of values, indicating both positive and negative possibilities or a margin of uncertainty; ↓: Signifies a decrease; ↑: Signifies an increase.*

## Methodology

This Review synthesises observational, satellite, and modelling evidence on the biophysical climate feedbacks of boreal afforestation, with emphasis on aerosol–cloud interactions and their implications for Earth's Energy Imbalance (EEI) [1, 3]. The spatial focus corresponds to the circumpolar boreal biome (Fig. S1), which spans North America and Eurasia and contains the primary flux tower, atmospheric monitoring, and long-term research stations that form the empirical foundation of the studies reviewed. These include the BOREAS and BERMS networks in Canada, AmeriFlux sites in Alaska, and comparable field observatories in Norway, Sweden, and Russia.

Relevant literature was identified through searches in Web of Science, Scopus, and Google Scholar using combinations of terms such as "boreal forest," "afforestation," "albedo," "snow masking," "biogenic volatile organic compounds," "secondary organic aerosols," "cloud condensation nuclei," and "radiative forcing." The scope was restricted to peer-reviewed studies and assessment reports published between 2000 and 2024 that quantified, modelled, or observed processes directly relevant to boreal forest–climate coupling [13, 16, 35, 106.]

Evidence was drawn from three complementary sources: (i) Ground-based observations, including eddy-covariance measurements of surface energy and water fluxes, aerosol concentrations, and micrometeorology [15, 49]; (ii) Satellite retrievals such as MODIS, CERES, and CALIPSO, which provide spatially resolved estimates of albedo, vegetation cover, and cloud properties [30, 16]; and (iii) Numerical simulations from global Earth System Models (e.g., NorESM, CMIP6-LUMIP), aerosol–climate models (e.g., GLOMAP), and process-based land surface models (e.g., CLASS-CTEM, CryoGrid) capable of representing coupled carbon, energy, and aerosol–cloud feedbacks [14,54].

Data integration followed a mechanistic framework, classifying results into three principal pathways: surface albedo changes [35, 112], surface energy and moisture exchange [15], and BVOC–aerosol–cloud processes [13, 47, 48]. Particular attention was given to studies that distinguished local climate responses from non-local, circulation-mediated effects of land-cover change [58, 59]. Reported radiative forcing values were retained as ranges to capture variability linked to species composition, stand age, disturbance history, and atmospheric background conditions.

Uncertainty assessment considered methodological issues (e.g., scale mismatches between site measurements and model grid cells), observational biases (notably over snow- and ice-covered surfaces [110]), and incomplete representation of BVOC chemistry, SOA yields, and CCN activation under cold, low-aerosol regimes [42, 50]. This integrative approach enabled a spatially explicit, process-oriented synthesis of the competing warming and cooling influences of boreal afforestation, situating snow–albedo dynamics and temperature-sensitive aerosol–cloud feedbacks at the centre of the policy-relevant analysis.

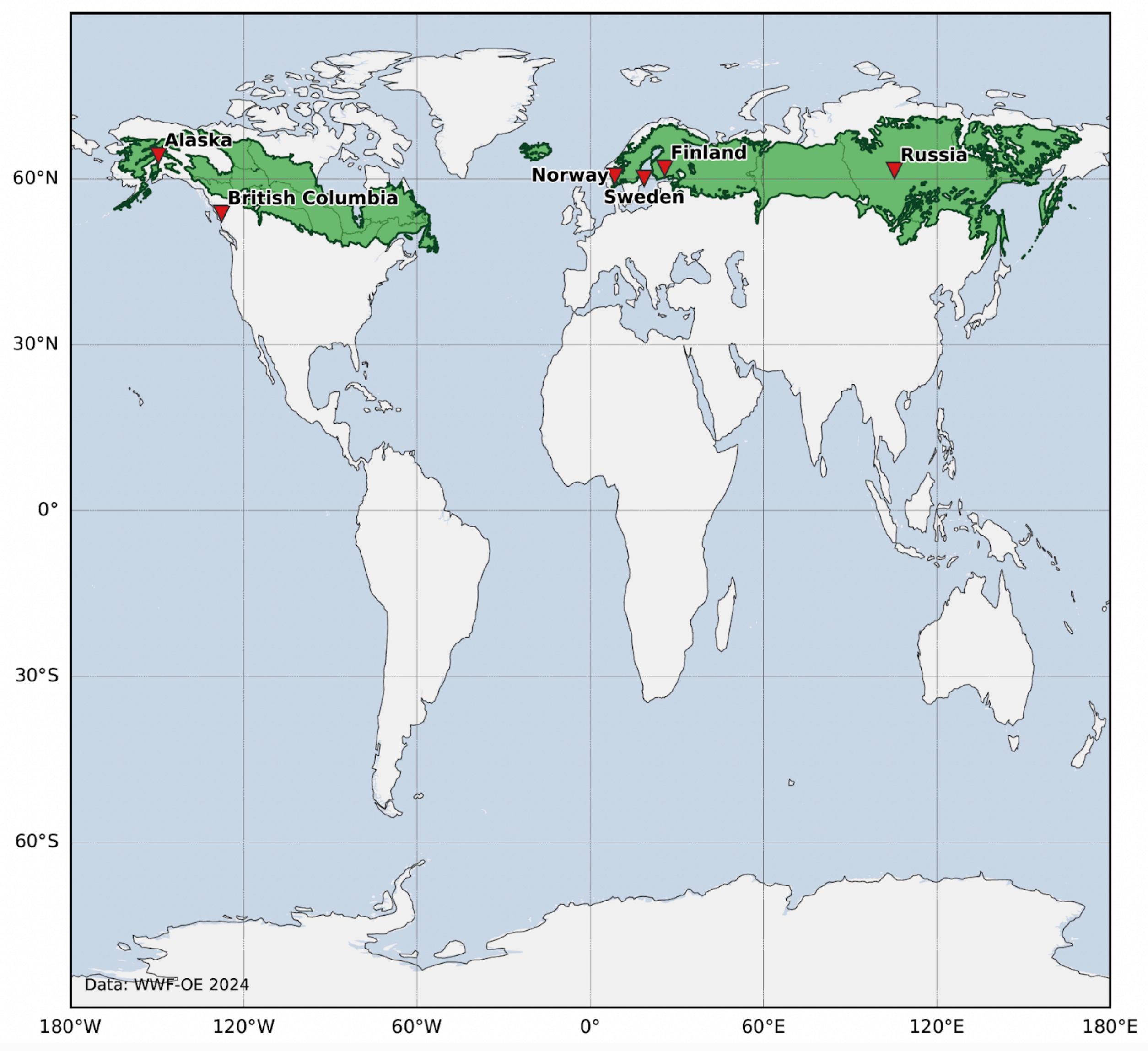

**Fig. S1 | Geographic distribution of the boreal forest biome and key data-source locations.** The map displays the circumpolar boreal forest biome (shaded in green; data: WWF-OE 2024). Red triangles indicate the general locations of key empirical data sources, flux tower networks (e.g., BOREAS/BERMS, AmeriFlux), and long-term research stations (e.g., SMEAR II, Hyytiälä) in Alaska, British Columbia, Norway, Sweden, Finland, and Russia. These sites provided the foundational data and model-validation points for the key studies on biophysical climate feedbacks that are synthesized and reviewed in this paper.